\documentclass[twocolumn,showpacs,prl,superscriptaddress,floatfix,amsmath,amssymb]{revtex4}

\usepackage{bm}
\usepackage{graphicx}

\DeclareMathOperator{\Tr}{Tr}
\newcommand{\parent}[1]{\left( #1\right)}
\newcommand{\abs}[1]{\left| #1\right|}
\newcommand{\bra}[1]{\left< #1\right|}
\newcommand{\ket}[1]{\left| #1\right>}
\newcommand{\bracket}[2]{\left< #1\: \right| \left. #2\right>}

\begin{document}

\title{Fidelity Decay for Phase Space Displacements}
\author{Diego V. Bevilaqua}
\affiliation{Department of Physics, Harvard University, Cambridge, MA 02138}
\author{Eric J. Heller}
\affiliation{Department of Physics, Harvard University, Cambridge, MA 02138}
\affiliation{Department of Chemistry and Chemical Biology, Harvard University, Cambridge, MA 02138}

\date{\today}

\begin{abstract}
In this letter we analyse the behavior of fidelity decay under a very specific kind of perturbation: phase space displacements. Under these perturbations, systems will decay following the Lyapunov regime only. Others universal regimes discussed in the literature are not presented in this case; instead, for small values of the perturbation we observe quantum freeze of the fidelity. We also show that it is possible to connect this result with the incoherent neutron scattering problem
\end{abstract}


\maketitle

Fidelity decay involves the overlap of two initially identical wavefunctions evolving under slightly different Hamiltonians~\cite{peres}.  This is also known as Loshmidt echo.  The subject has received much attention over the past few years as a model useful for discussion of decoherence and/or quantum classical correspondence~\cite{jalabert,vanicek,lyapunov-measure,tomsovic}. For a generic perturbation of the Hamiltonian, it was shown that the decay follows three different universal regimes~\cite{vanicek}: the Gaussian regime, the Fermi Golden Rule (FGR) regime and the Lyapunov regime, followed by one strong non-generic regime. The Lyapunov regime is especially interesting because it is a  way to measure the \textit{classical} Lyapunov exponent of a system from a pure \textit{quantum} measurement~\cite{lyapunov-measure}; and it is a  universal environment-free coherence decay~\cite{zurek}.

In this letter we analyze the behavior of Loshmidt Echo under a non-generic perturbation of the Hamiltonian, namely phase space displacements. These Loshmidt types are suggested by experiments involving neutron scattering (where the phase space displacement is a momentum boost~\cite{ins}) and to some extent by electronic transitions in molecules and solids, where the displacement is along the position axis, with very little other change in the potential~\cite{displacement-spectroscopy}. The non-generic character of this kind of perturbation might not be expected to follow the usual sequence of regimes, and indeed we show the phase space displacement presents only one of the universal regimes, the Lyapunov one.  In this letter we will focus on  incoherent neutron scattering. In general any kind of experimental measurement of momentum-momentum time correlation, position-position time correlation or a combination of both can be viewed as a Loshmidt Echo under phase space displacements.

Under the formalism as presented by Lovesey~\cite{lovesey}(which is based on Van Hove's linear response theory~\cite{vanhove}), the incoherent part of differential cross section for neutron scattering can be calculated from the correlation function,
\[
Y_{jj}\parent{\bf{k},t}=\left<  e^{-i\bf{k}\cdot \bf{\hat{r}}_j}e^{\frac{i\hat{H}}{\hbar}t}e^{i\bf{k}\cdot \bf{\hat{r}}_j}e^{\frac{-i\hat{H}}{\hbar}t}\right>,
\]
where the brackets means ensemble average, $\bf{\hat{r}}_j$ are the position operators of the nuclei and $\hat{H}$ is the typical Hamiltonian of the target system. In the Heisenberg representation, the correlation function can be written as:
\[
Y_{jj}\parent{\bf{k},t}=\left< e^{-i\bf{k}\cdot \hat{\bf{r}}_j\parent{0}}e^{i\bf{k}\cdot \hat{\bf{r}}_j\parent{t}}\right>.
\]
The ensemble average of the correlation function can be approximated by~\cite{ins}:
\begin{equation}
\begin{split}
Y_{jj}\parent{\bf{k},t}& \approx \frac{1}{Q}\int \parent{\frac{d^{2N}\alpha}{\pi^N}}\Phi\parent{\alpha} \times \\
& \times \bra{\alpha}e^{-i\bf{k}\cdot \hat{\bf{r}}_j}e^{\frac{i\hat{H}}{\hbar}t}e^{i\bf{k}\cdot \hat{\bf{r}}_j}e^{\frac{-i\hat{H}}{\hbar}t}\ket{\alpha}, \label{auto}
\end{split}
\end{equation}
where $\ket{\alpha}$ are coherent states in $N$ degrees of freedom, $Q=\Tr \left[ e^{-\beta \hat{H}}\right]$, and $\Phi\parent{\alpha}=e^{-\beta H_{cl}\parent{\alpha}}$. Introducing the notation $e^{\frac{i\hat{H}_{\bf{k}}}{\hbar}t}=e^{-i\bf{k}\cdot \hat{\bf{r}}_j}e^{\frac{i\hat{H}}{\hbar}t}e^{i\bf{k}\cdot \hat{\bf{r}}_j}$, we can identified the kernel of the integral $I\parent{t}=\bra{\alpha}e^{-i\bf{k}\cdot \hat{\bf{r}}_j}e^{\frac{i\hat{H}}{\hbar}t}e^{i\bf{k}\cdot \hat{\bf{r}}_j}e^{\frac{-i\hat{H}}{\hbar}t}\ket{\alpha}$ as the same kernel of a typical Loshmidt Echo problem:
\begin{equation}
M\parent{t}=\abs{\bra{\alpha}e^{\frac{i\hat{H}_{\bf{k}}}{\hbar}t} e^{\frac{-i\hat{H}}{\hbar}t}\ket{\alpha}}^2=\abs{I\parent{t}}^2.
\label{fidelitydecay}
\end{equation}

\begin{figure}
\begin{center}
\fbox{\includegraphics[width=7.0cm]{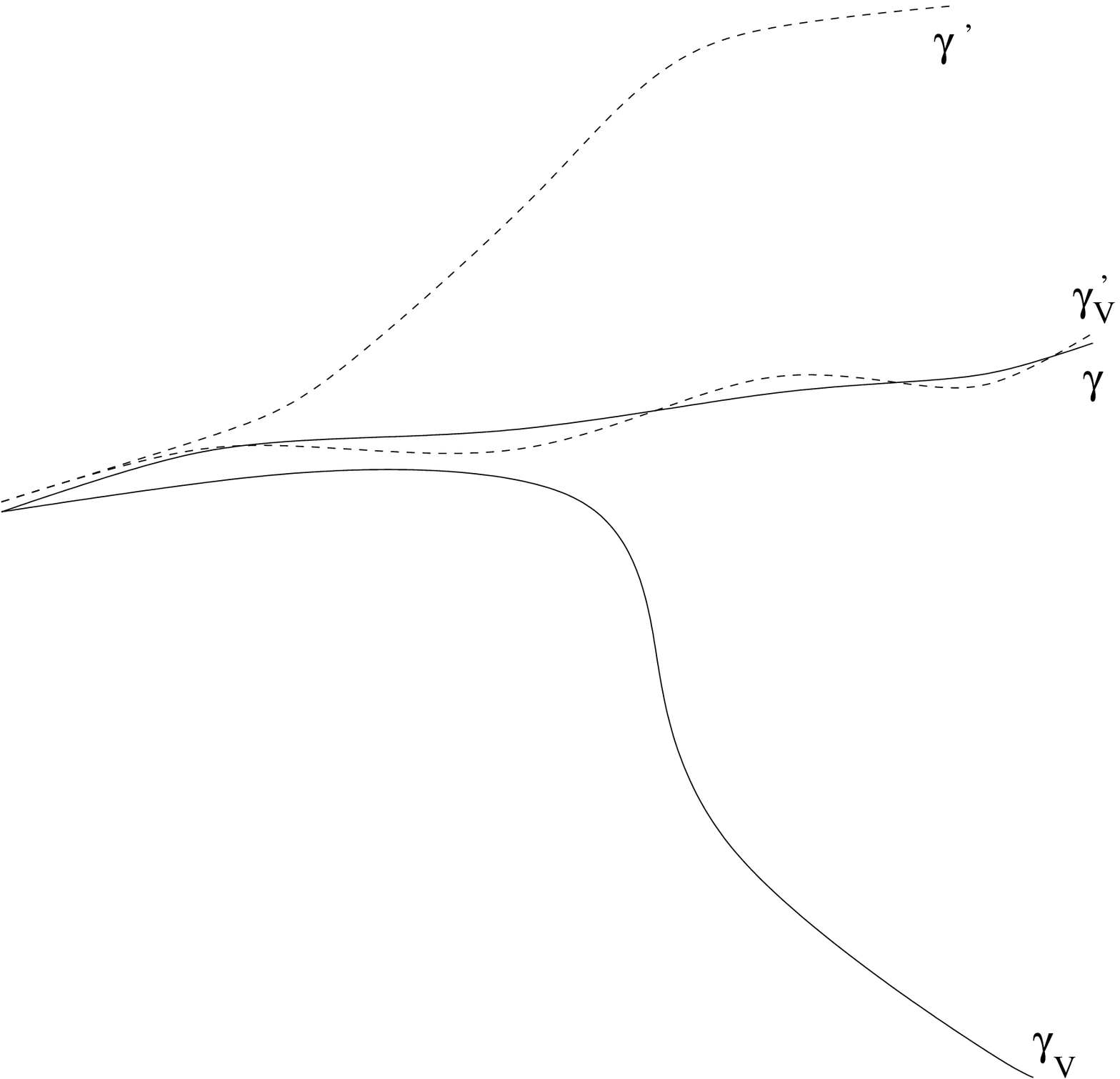}}
\vspace{0.8cm} \\
\fbox{\includegraphics[width=7.0cm]{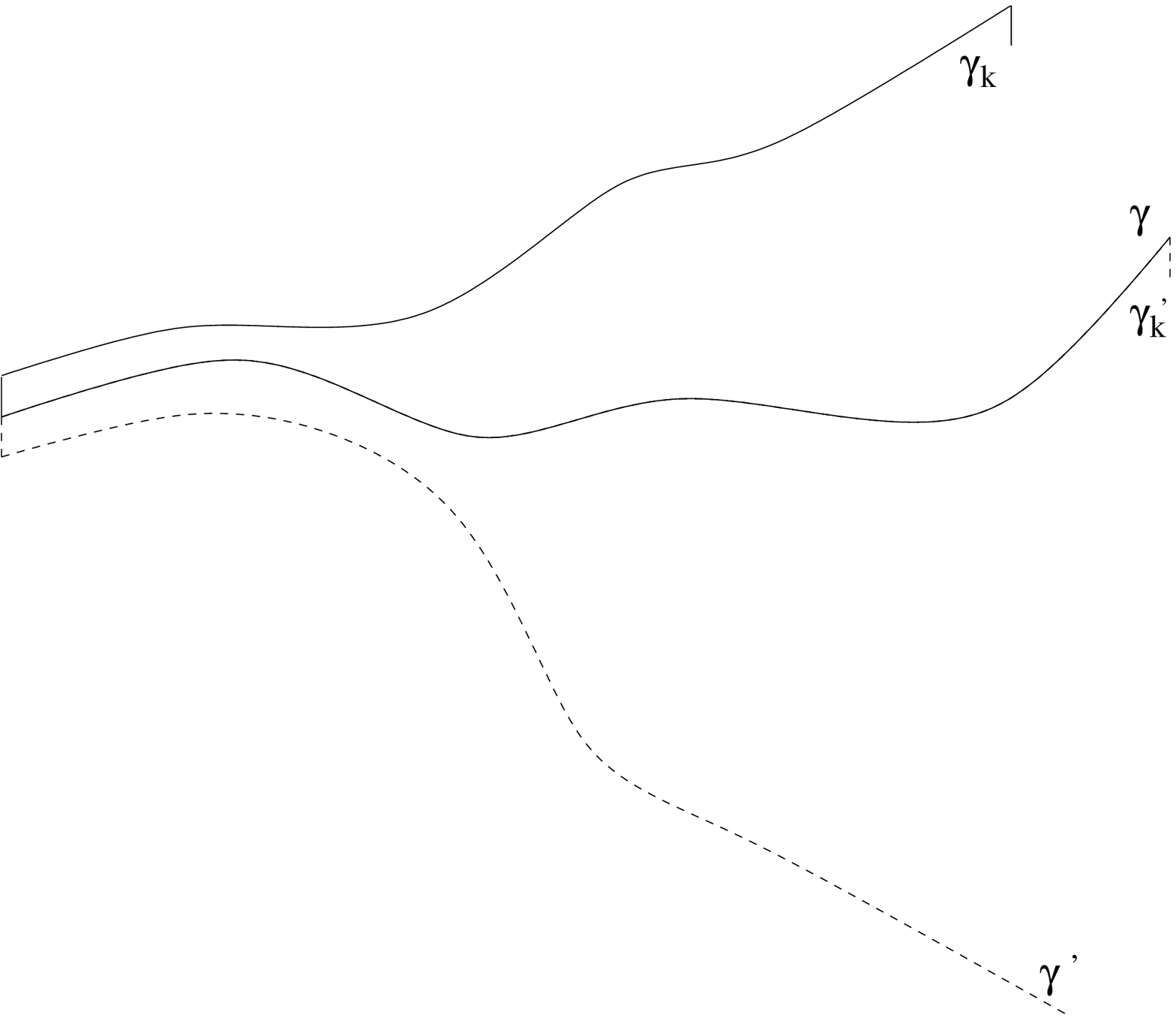}}
\end{center}
\caption{Illustrative view of structural stability or orbits. Top plot: Structural stability for generic perturbations; $\gamma$ is a generic orbit of unperturbed Hamiltonian, $\gamma_V$ is the same initial condition under perturbed Hamiltonian, $\gamma^{\prime}$ is a different orbit of the unperturbed Hamiltonian, and $\gamma_V^{\prime}$ is this different orbit under perturbation. Bottom plot: Phase space displacements; Labels are the same, except that subscript $\bf{k}$ is used instead of $V$ for perturbed orbits. Notice that $\gamma_{\bf{k}}^{\prime}$ is barely visible since it is hidden under orbit $\gamma$.}\label{structural_figure}
\end{figure}

Before we carry out a semiclassical evaluation of the fidelity decay in phase space displacement problems, it is interesting to analyze why it is possible to use the \textit{Diagonal Approximation} as called in~\cite{jalabert}, and why this approximation is even better in the case analyzed in this letter.

The \textit{Diagonal Approximation} is based on the assumption that, given an perturbed Hamiltonian $H_V=H+V$, one can, classically, make the approximation that a trajectory $\gamma\parent{t}$ in configuration space does not change, though its action is changed by $\Delta S^{\gamma}=S_V^{\gamma}-S^{\gamma}=\int_0^t dt' V\left[ \gamma\parent{t'}\right]$. Unfortunately this assumption is not valid anymore for chaotic system due to the exponential divergence of single orbits.

As pointed out by Cerruti and Tomsovic~\cite{tomsovic}, and Vani\v{c}ek and Heller~\cite{vanicek} the reason this approximation is still valid for chaotic systems under a semiclassical formalism is mainly due to structural stability theorems~\cite{sst}. Given a uniformly hyperbolic Hamiltonian system, under a generic small perturbation it is almost always possible to find a trajectory $\gamma_V^{\prime}\parent{t}$ of the perturbed system which is arbitrarily close to the some unperturbed trajectory $\gamma\parent{t}$ (even though they do not share common end points). Since  semiclassical integrals involve a set of neighboring trajectories, both $\gamma_V^{\prime}\parent{t}$ and $\gamma\parent{t}$ are included. In the classical limit of the Loshmidt Echo, this is equivalent to saying that the particle follows $\gamma\parent{t}$ in the forward direction, but $\gamma_V^{\prime}\parent{t}$ in the backwards direction (i.e., that is the path which minimizes the action!).

If the perturbation is given by a uniform boost in phase space, the diagonal approximation is \textit{exact} apart from a time independent correction to the action: For any given trajectory $\gamma\parent{t}$, with end points $\parent{\bf{q}_0,\bf{q}_f}$, there is a trajectory of the perturbed Hamiltonian $H_{\bf{k}}$, $\gamma_{\bf{k}}^{\prime}\parent{t}$ with end points $\parent{\bf{q}_0^{\prime},\bf{q}_f^{\prime}}$ that follows exactly the previous path, aside from small corrections at the extremities. The correction to the action is small and time independent. The fact that each individual \textit{pseudo-trajectory}~\cite{pseudo-trajectory} $\gamma\parent{t}+\gamma_V^{\prime}\parent{t}$ has a time \textit{dependent} correction to the action is fundamental to the existence of the FGR regime in generic Loshmidt Echo system.  This is an intuitive reason why the FGR regime is absent in the case of phase space displacements. In Fig.~\ref{structural_figure} this discussion is illustrated.

For a more quantitative approach to the problem, we want to semiclassically evaluate
\begin{equation*}
\begin{split}
M\parent{t}& =\abs{\bra{\alpha}e^{-i\mathbf{k}\cdot \hat{\mathbf{r}}}e^{\frac{i\hat{H}}{\hbar}t}e^{i\mathbf{k}\cdot \hat{\mathbf{r}}}e^{-\frac{i\hat{H}}{\hbar}t}\ket{\alpha}}^2 \\
& =\abs{I\parent{t}}^2
\end{split}
\end{equation*}
where $\ket{\alpha}$ is a coherent Gaussian wavepacket,
\[
\bracket{\mathbf{r}}{\alpha \parent{\mathbf{r}_0,\mathbf{q}_0}}=\parent{\pi \sigma^2}^{-\frac{d}{4}}e^{\frac{i}{\hbar} \mathbf{p}_0 \cdot \parent{\mathbf{r}-\mathbf{r}_0}-\frac{\parent{\mathbf{r}-\mathbf{r}_0}^2}{2 \sigma^2}}.
\]
Following~\cite{heller-propagator}, we can approximate the semiclassical propagation of a coherent Gaussian wave packet using the Van Vleck propagator and expanding it around $\bf{r}_0$:
\begin{equation*}
\begin{split}
\bra{\mathbf{r'}}e^{-\frac{i}{\hbar}\hat{H}t}\ket{\alpha}_{SC} & \approx
\parent{-\frac{i\sigma}{\sqrt{\pi}\hbar}}^{\frac{d}{2}} \times \\
& \times \sum_j \sqrt{C_j} e^{i\frac{S_j}{\hbar}-i\frac{\pi \nu_j}{2}}e^{-\frac{\sigma^2}{2\hbar^2}\parent{\mathbf{p}-\mathbf{p}_0}},
\end{split}
\end{equation*}
where $\mathbf{p}=\left. -\partial S / \partial \mathbf{r}\right|_{\mathbf{r}=\mathbf{r}_0}$, the classical momentum for a trajectory starting on $\mathbf{r}_0$ and finishing on $\mathbf{r}'$ , $S_j$ is the classical action along the same path, $\nu_j$ the Maslov index and $C_j=\abs{-\partial^2 S_j\parent{\mathbf{r},\mathbf{r}';t} / \partial r_i\partial r^{\prime}_j}_{\mathbf{r}=\mathbf{r}_0}$, all calculated over a given allowed classical trajectory $\gamma_j$ and summed over all possible classical trajectories.

The kernel $I\parent{t}$ involves a double sum over all classical trajectories, $j$ and $j'$, such that it can be interpreted as a overlap between two wavepackets, one that is boosted and subsequently propagated and another one that is propagated and subsequently  boosted~\cite{ins}. However, as pointed out before, one can assume that only the overlap between the two trajectories that minimize the action of the pseudo-trajectory contributes significantly, while the others oscillate strongly. The two trajectories have the same action and the same Maslov index, so we can just assume that $j=j'$ and write:
\begin{equation*}
\begin{split}
I\parent{t}& =\parent{-\frac{\sigma^2}{\pi \hbar^2}}^{\frac{d}{2}}e^{-i\mathbf{k} \cdot \mathbf{r}_0} \times \\
& \times \int d^dr' \sum_j C_j e^{i\mathbf{k} \cdot \mathbf{r}'}e^{-\frac{\sigma^2}{2\hbar^2}\left[ \parent{\mathbf{p}_j-\mathbf{p}_0}^2+\parent{\mathbf{p}_j-\mathbf{p}_0-\hbar \mathbf{k}}^2 \right]}.
\end{split}
\end{equation*}
The last equation involves finding all possible classical paths which return to their initial position after time $t$. These can be   large in number, and worse, the prefactors $C_j$ can diverge. Vani\v{c}ek and Heller in~\cite{vanicek} introduce the idea of using the IVR representation~\cite{ivr} in a similar integral. Because there are actually two semiclassical integrals simplified in the single integral above, instead of the usual term $\sqrt{C_j}$, we have $C_j$. However the term $C_j$ is Jacobian of the transformation and disappear after the change from final position representation to initial momentum representation. After transformation is done, we do not only eliminate the divergences in the integral, but we do not even need to calculate the term $C_j$ anymore. We do not need also to consider more that one path, since the integral in initial momentum representation has a unique allowed classical path. Introducing a new variable $\boldsymbol{\xi}=\parent{\mathbf{p}-\mathbf{p}_0} / \hbar-\mathbf{k} / 2$, the integral becomes, after a change to initial momentum representation:
\[
I\parent{t}=\parent{-\frac{\sigma^2}{\pi}}^{\frac{d}{2}}e^{-i\mathbf{k} \cdot \mathbf{r}_0}e^{-\frac{\parent{\sigma \mathbf{k}}^2}{4}}
\int d^d\xi e^{i\mathbf{k}\cdot \mathbf{r}'}e^{-\sigma^2 \boldsymbol{\xi}^2},
\]
where $\mathbf{r}' =\mathbf{r}'(\mathbf{r}_0, \mathbf{p}\parent{\mathbf{p}_0,\boldsymbol{\xi}};t)$ is the final position for a particle with initial position $\mathbf{r}_0$ and initial momentum $\mathbf{p}$.

$M\parent{t}$ is then given by:
\begin{equation*}
\begin{split}
M\parent{t}& =\parent{\frac{\sigma^2}{\pi}}^d e^{-\frac{\parent{\sigma \mathbf{k}}^2}{2}} \times \\
& \times \iint d^d\xi_1 d^d\xi_2 e^{i\mathbf{k}\cdot \parent{\mathbf{r'_1-r'_2}}}e^{-\sigma^2 \parent{\boldsymbol{\xi}_1^2+\boldsymbol{\xi}_2^2}}.
\end{split}
\end{equation*}
For a chaotic system, trajectories with a small initial separation $\Delta \mathbf{q}_0$ in phase space, i.e.  $\Delta \mathbf{q}_0 \ll \Delta L e^{-\lambda}$ (where $\Delta L$ is the size of phase space and $\lambda$ the leading classical Lyapunov exponent of the system) we have that:
\[
\abs{\mathbf{r}'_1-\mathbf{r}'_2} \sim \Delta \mathbf{q} \sim \frac{\sigma^2\abs{\mathbf{p}_1-\mathbf{p}_2}}{\hbar}e^{\lambda t}=\sigma^2 \abs{\boldsymbol{\xi}_1-\boldsymbol{\xi}_2}e^{\lambda t}.
\]
On the other hand, if $\Delta \mathbf{q}_0 > \Delta L e^{-\lambda}$, then trajectories are uncorrelated and the mean separation is not dependent on time. In general, $M\parent{t}$ comes in two pieces, correlated and uncorrelated. The uncorrelated part $M_u \parent{t}$ can be calculated assuming a random phase approximation in the integral above,
\[
M_u\parent{t}=e^{-\frac{\parent{\sigma \mathbf{k}}^2}{2}},
\]
that is, the uncorrelated part is a small time independent decay. In the generic Loshmidt Echo problem, the decay due to the uncorrelated trajectories is recognized as the responsible for the FGR part of the fidelity decay~\cite{vanicek}. The result above means that \textit{there is no FGR regime in the fidelity decay of phase space displacements}.

Before we calculate the correlated part of the decay, we should notice that there is also no Gaussian decay in this problem. The Gaussian decay in generic Loshmidt Echo problems is obtained through first order quantum mechanics perturbation theory. However, the displacements in phase space do not change the spectrum of the system (besides some possible shift of the ground state), so all the first order correction are zero, as the Gaussian contribution to the fidelity decay. This is actually analogous to the problem of spectrum change due to perturbation in the shape of billiards~\cite{doron-alex}. When the billiards are simply displaced, there is decay of the overlap of eigenfunctions for example but  no change in the spectrum.

In order to calculate the correlated part of the fidelity decay, we will approximate the term $\exp \parent{i\mathbf{k}\cdot\parent{\mathbf{r}^{\prime}_1-\mathbf{r}^{\prime}_2}}$ by its average over phase space. Because of the symmetry of phase space, $\mathbf{r}^{\prime}_1-\mathbf{r}^{\prime}_2$ will approach a Gaussian centered on zero with variance:
\[
\left< \parent{\mathbf{r}^{\prime}_1-\mathbf{r}^{\prime}_2}^2\right> \sim \sigma^4 \parent{\boldsymbol{\xi}_1-\boldsymbol{\xi}_2}^2e^{2\lambda t}.
\]
So,
\[
e^{i\mathbf{k}\cdot\parent{\mathbf{r}^{\prime}_1-\mathbf{r}^{\prime}_2}} \approx e^{-\mathbf{k}^2\sigma^4 \parent{\boldsymbol{\xi}_1-\boldsymbol{\xi}_2}^2e^{2\lambda t}},
\]
and the correlated part of fidelity decay become:
\begin{equation*}
\begin{split}
M_c\parent{t} & \approx \parent{\frac{\sigma^2}{\pi}}^d e^{-\frac{\parent{\sigma \mathbf{k}}^2}{2}} \times \\
& \times \iint d^d\xi_1 d^d\xi_2 e^{-\mathbf{k}^2\sigma^4 \parent{\boldsymbol{\xi}_1-\boldsymbol{\xi}_2}^2e^{2\lambda t}}e^{-\sigma^2 \parent{\boldsymbol{\xi}_1^2+\boldsymbol{\xi}_2^2}} \\
& \approx e^{-\frac{\parent{\sigma \mathbf{k}}^2}{2}} \parent{1+2\parent{\mathbf{k}\sigma}^2e^{2\lambda t}}^{\frac 12} \\
& \sim e^{-\frac{\parent{\sigma \mathbf{k}}^2}{2}} \frac{e^{-\lambda t}}{\sqrt{2}\abs{\mathbf{k}}\sigma}.
\end{split}
\end{equation*}
The last line reveals that the Lyapunov decay still existing for the phase space displacement and it is actually the only time dependent fidelity decay for this kind of perturbation. The only condition we required upon the systems is the classical uniformly hyperbolicity. This is the first example of a pure Lyapunov decay of fidelity; it is also experimentally realizable. Because the fidelity decay $M\parent{t}$ is already averaged over coherent states, its thermal average is itself, $\left< M\parent{t}\right>$. So, it is straight forward to show that the autocorrelated function as defined in~\eqref{auto} is just:
\[
Y_{jj}\parent{\bf{k},t} \approx\abs{\left< I\parent{t}\right>} = \sqrt{M\parent{t}}.
\]

\begin{figure}
\begin{center}
\includegraphics[width=7.5cm]{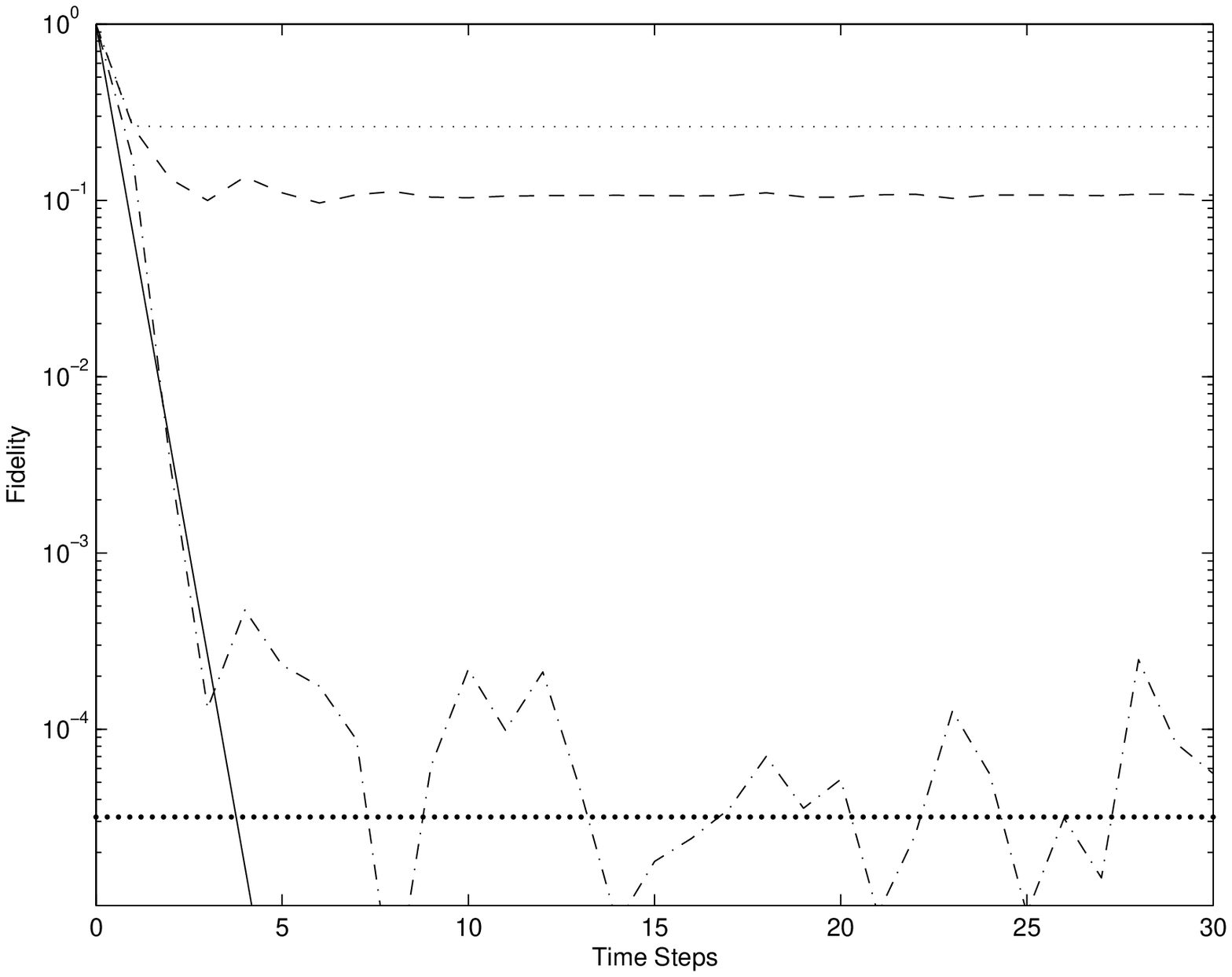}
\vspace{0.8cm} \\
\includegraphics[width=7.5cm]{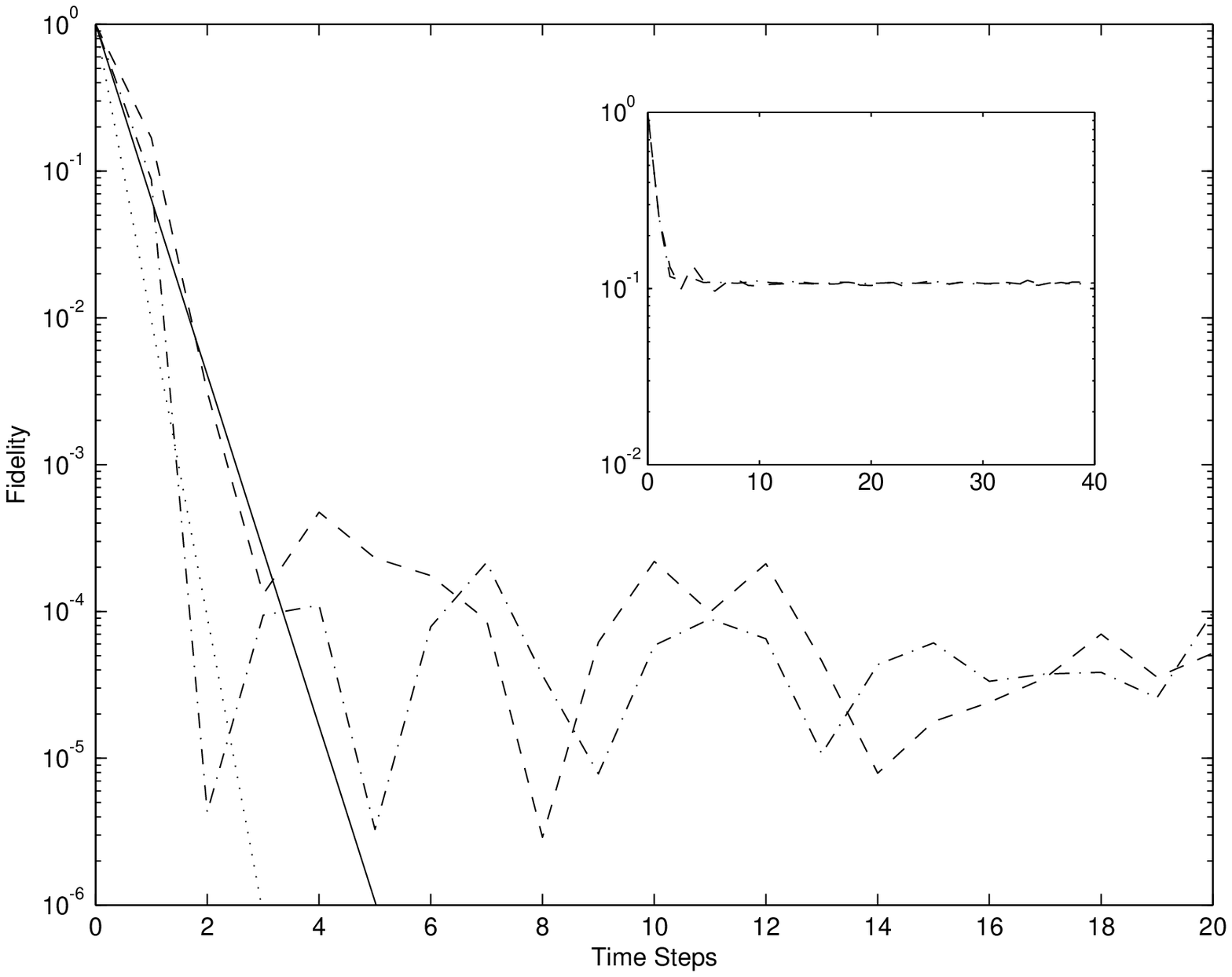}
\end{center}
\caption{Top Plot: Fidelity decay for different perturbation values. Dotted, dashed and dot-dashed lines are numerically calculated for the standard map with boost $k = 10^{-5}, 10^{-4},10^{-3}$, respectively. Solid line is exponential decay at Lyapunov rate (where $\lambda = 2.7505$) and ``big''-dotted line is the ``quantum'' ergodic limit. All numerical calculations where done with kick strength $K_s = 7.0$, and averaged over 20 coherent states. Bottom Plot: Fidelity decay for different kick strengths. Solid and dotted lines are Lyapunov decays with $\lambda = 2.7505$ and $4.6497$, respectively ($K_s$ of $7.0$ and $18.0$). The dashed and dot-dashed lines are numerically calculated with $K_s$ of $7.0$ and $18.0$, respectively, and both with $k = 10^{-3}$. Inset on bottom plot shows the numerical calculated fidelity decay for $k = 10^{-4}$, with dashed line correspoding to $K_s = 7.0$ and dot-dashed corresponding to $K_s = 18.0$} \label{fig-fidelity-decay}
\end{figure}

As a check and illustration of this work, we numerically calculated the fidelity decay for a quantum standard map with phase space displacement perturbations to compare with our predictions made with the semiclassical formalism. The results are shown in figure~\ref{fig-fidelity-decay} for different values of momentum boost $k$ and different values of kick strength of the map $K_s$ (which means different Lyapunov exponents). All numerical calculations were made with $N=5000$ states. For small values of the ``perturbation strength'' ($k$), we notice that the time-dependent behavior of the decay is suppressed and the fidelity is frozen in a `plateau' that is independent of the parameters of the map ($K_s$)~\cite{prosen}. For bigger values of $k$, the results seen to follow a Lyapunov decay until it reaches the quantum ergodic limit. To be more precise, given a quantization of the momentum and mean separation between momentum levels $\overline{\Delta p}$, if the perturbation $k >> \overline{\Delta p}$ then the Lyapunov decay is observed. On the other side, if $k << \overline{\Delta p}$ then the fidelity is frozen on plateau with value $\exp \parent{-\frac{\parent{k\sigma}^2}{2}}$. In between, there is a transition region where the fidelity is frozen at intermediate values.

In conclusion, we showed that interesting things happen when we focus on particular form of perturbation in Loshmidt Echo problems. In particular, if we choose phase space displacement as particular form for the perturbation, all the time dependent contribution to the fidelity decay vanishes beside the Lyapunov contribution. This is not only an interesting example but also a naturally arising perturbation in the incoherent part of the differential cross section in neutron scattering.

D. V. Bevilaqua was partially supported from CAPES, Coordena\c{c}\~{a}o de Aperfei\c{c}oamento de Pessoal de N\'{i}vel Superior and NSF grant no. CHE-0073544.


\begin{thebibliography}{99}
\bibitem{pseudo-trajectory} By pseudo-trajectory we mean a union of the trajectories $\gamma\parent{t'}$ and $\gamma \parent{t+t'}$. In a Loshmidt Echo language, that is the path which minimizes the classical action of a system that goes forward with Hamiltonian $H$ and backwards with Hamiltonian $H^V$.
\end{thebibliography}
\end{document}